\documentclass[prl,twocolumn,superscriptaddress]{revtex4-1}
\usepackage{graphicx}
\usepackage{dcolumn}
\usepackage{bm}
\usepackage{color}
\usepackage{natbib}

\begin{document}
\title {Quantum interference in an interfacial superconductor}
\author{Srijit Goswami}
\altaffiliation{Equal contributions}
\affiliation{Kavli Institute of Nanoscience, Delft University of Technology, P.O. Box 5046, 2600 GA Delft, The Netherlands.}
\author{Emre Mulazimoglu}
\altaffiliation{Equal contributions}
\affiliation{Kavli Institute of Nanoscience, Delft University of Technology, P.O. Box 5046, 2600 GA Delft, The Netherlands.}
\author{Ana M. R. V. L. Monteiro}
\affiliation{Kavli Institute of Nanoscience, Delft University of Technology, P.O. Box 5046, 2600 GA Delft, The Netherlands.}
\author{Roman W\"{o}lbing}
\affiliation{Physikalisches Institut and CQ Center for Quantum Science in LISA$^+$, Eberhard Karls Universit\"{a}t T\"{u}bingen, Auf der Morgenstelle 14, D-72076, T\"{u}bingen, Germany.}
\author{Dieter Koelle}
\affiliation{Physikalisches Institut and CQ Center for Quantum Science in LISA$^+$, Eberhard Karls Universit\"{a}t T\"{u}bingen, Auf der Morgenstelle 14, D-72076, T\"{u}bingen, Germany.}
\author{Reinhold Kleiner}
\affiliation{Physikalisches Institut and CQ Center for Quantum Science in LISA$^+$, Eberhard Karls Universit\"{a}t T\"{u}bingen, Auf der Morgenstelle 14, D-72076, T\"{u}bingen, Germany.}
\author{Ya. M. Blanter}
\affiliation{Kavli Institute of Nanoscience, Delft University of Technology, P.O. Box 5046, 2600 GA Delft, The Netherlands.}
\author{Lieven M. K. Vandersypen}
\affiliation{Kavli Institute of Nanoscience, Delft University of Technology, P.O. Box 5046, 2600 GA Delft, The Netherlands.}
\author{Andrea D. Caviglia}
\affiliation{Kavli Institute of Nanoscience, Delft University of Technology, P.O. Box 5046, 2600 GA Delft, The Netherlands.}

\maketitle

\textbf{The two-dimensional superconductor formed at the interface between the complex oxides, lanthanum aluminate (LAO) and strontium titanate (STO)~\cite{Reyren_Science} has several intriguing properties~\cite{Caviglia_SO_PRL,Shalom_SO_PRL,Dikin_Coex_PRL,Bert_Moler_Nat_Phys,Li_Ashoori_NatPhys} that set it apart from conventional superconductors. Most notably, an electric field can be used to tune its critical temperature (T$_c$)~\cite{Caviglia_Nature}, revealing a dome-shaped phase diagram reminiscent of high T$_c$ superconductors~\cite{Richter_Nature_Gap}. So far, experiments with oxide interfaces have measured quantities which probe only the magnitude of the superconducting order parameter and are not sensitive to its phase. Here, we perform phase-sensitive measurements by realizing the first superconducting quantum interference devices (SQUIDs) at the LAO/STO interface. Furthermore, we develop a new paradigm for the creation of superconducting circuit elements, where local gates enable in-situ creation and control of Josephson junctions.  These gate-defined SQUIDs are unique in that the entire device is made from a single superconductor with purely electrostatic interfaces between the superconducting reservoir and the weak link. We complement our experiments with numerical simulations and show that the low superfluid density of this interfacial superconductor results in a large, gate-controllable kinetic inductance of the SQUID. Our observation of robust quantum interference opens up a new pathway to understand the nature of superconductivity at oxide interfaces.}

A superconducting quantum interference device (SQUID) consists of two Josephson junctions (JJs) embedded in a superconducting loop. When a magnetic flux ($\Phi$) threads through this loop, it changes the relative difference in the superconducting phase of the two JJs giving rise to periodic oscillations in the supercurrent. This basic principle has been used with great success to study a variety of material systems. For example, standard superconductors have been combined with other materials such as ferromagnets~\cite{Ferro_AartsPRL_2001, Ferro_WeidesPRL_2012}, topological insulators~\cite{Veldhorst_Nature_BiSe} and nanowires~\cite{Kouwenhoven_SupCurrRev_Nature} to often reveal non-trivial current-phase relations. Such phase sensitive measurements have also emerged as a powerful tool to study more exotic superconductors such as the ruthenates~\cite{Ruthenate_SQUID_Science} and high $T_c$ cuprates~\cite{Harlingen_RMP}. Whether the two-dimensional (2D) superconductor formed at the LAO/STO interface is also unconventional is still not clear. However, to address this issue one must move beyond standard bulk transport measurements. Recent tunneling studies~\cite{Richter_Nature_Gap} and transport spectroscopy of confined structures~\cite{Levy_Nature_Gap} exemplify this point. In this context, a direct probe of the superconducting phase could provide complementary information about the microscopic origin of the superconductivity, but such experiments are missing. Here we realize SQUIDs at the LAO/STO interface, thereby enabling the first phase-sensitive measurements of this interfacial superconductor.

\begin{figure*}[!t]
\includegraphics[width=1\linewidth]{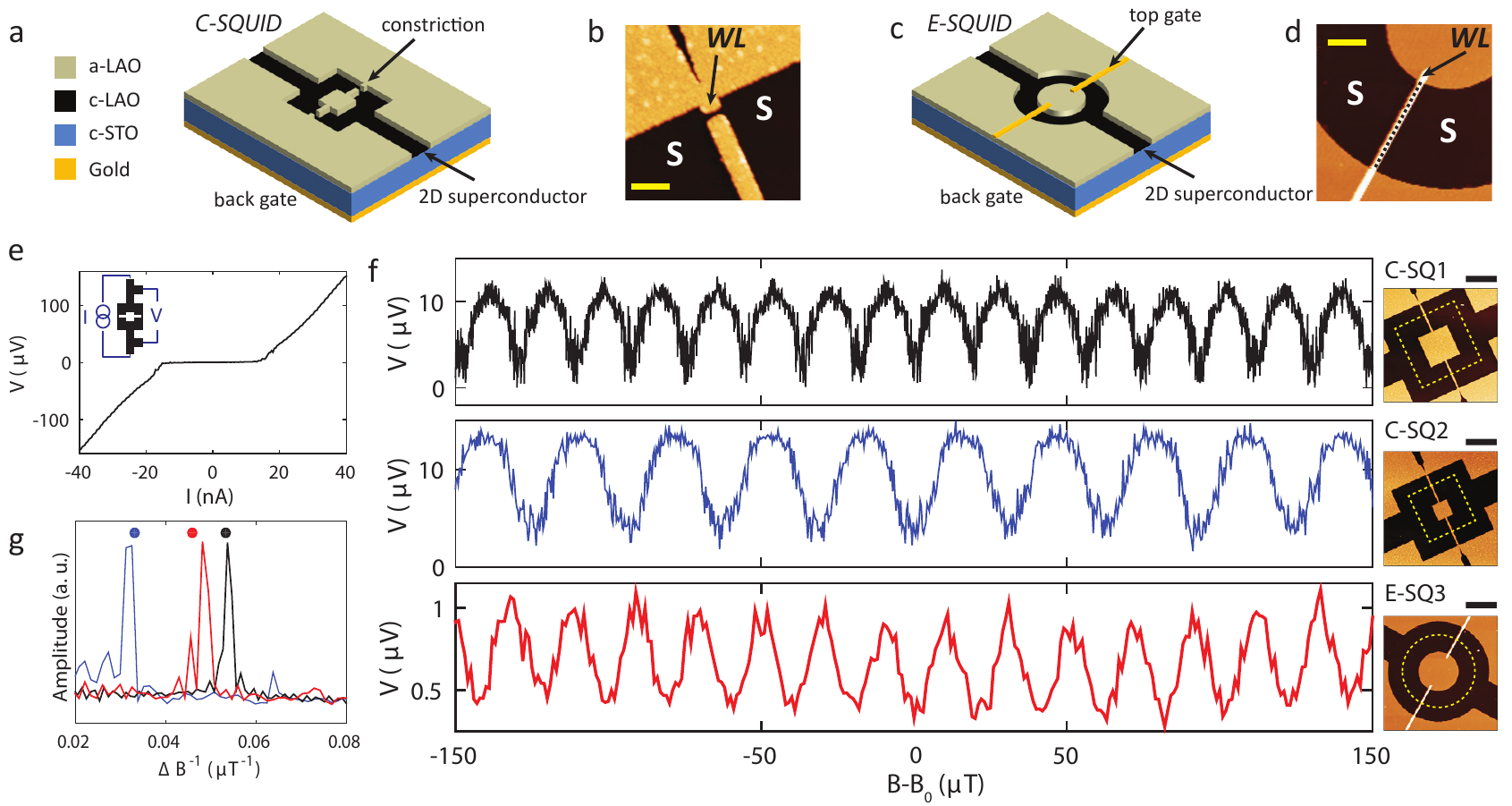}
\caption{\textbf{Device description and $V(\Phi)$ oscillations} (a) Schematic of constriction based SQUID (C-SQUID). (b) AFM image of the constriction, which serves as the weak link (WL). Scale bar is 400~nm. (c) Schematic of electrostatically defined SQUID (E-SQUID) (d) AFM image of region around the top gate (scale bar is 2~$\mu$m). Dashed line shows the region below the gate where the WL forms when a negative gate voltage is applied to the gate. (e) V-I curve for device C-SQ1 at $V_{bg}=-7$~V. Inset shows a schematic of the measurement configuration. (f) Oscillations in voltage ($V$) with magnetic field ($B-B_0$) for C-SQ1 (upper panel), C-SQ2 (middle panel) and E-SQ3 (lower panel) with AFM images of the respective devices. $B_0$ is an experimentally determined offset in the magnetic field and has an uncertainty greater than one oscillation period. Yellow dashed line marks the effective area threaded by the flux. Scale bar for all images is 5~$\mu$m. (g) Fourier transform of the oscillations shown in (f). The color scheme is the same as in (f). Circles are results of numerical simulations which yield an oscillation period in close agreement with experiments.}
\label{Fig1}
\end{figure*}

We fabricate the SQUIDs using two distinct approaches (see SI for full details). The first involves the creation of weak links using nanoscale physical constrictions (C-SQUIDs), a technique which has been used extensively in a wide variety of superconductors. Fig.~\ref{Fig1}a shows a schematic of the C-SQUID. Black areas are superconducting, while the beige regions remain insulating due to the presence of an amorphous LAO (a-LAO) mask. Each arm of the loop is interrupted by a narrow constriction (see AFM image in Fig.~\ref{Fig1}b). We ensure that the width of the constriction (sub-100~nm) is less than/comparable to the superconducting coherence length of LAO/STO~\cite{Reyren_Science}. The C-SQUID has the advantage that it requires only a single lithography step and is particularly convenient to characterize.

The second approach used to define SQUIDs, though more involved, is novel and unique to the LAO/STO interface. We exploit the sensitivity of $T_c$ to the field effect to create an electrostatically defined SQUID (E-SQUID). By applying negative voltages to local top gates~(see schematic in Fig.~\ref{Fig1}c), we deplete the regions below them. These locally depleted regions serve as the weak links, thus enabling the formation of independently tunable JJs in each arm. Fig.~\ref{Fig1}d shows an AFM image of one such gate-defined JJ. While other examples of gate-tunable JJs do exist~\cite{GateTune_InAs_PRL,GateTune_Nanowire_Science,Heersche_GrapheneJJ_Nature}, they necessarily involve physical interfaces between two dissimilar materials. In contrast, LAO/STO provides a unique material platform where a single superconductor can be electrostatically modified to allow in-situ creation and tuning of JJs in a perfectly reversible manner. In this work we study two C-SQUID devices (C-SQ1 and C-SQ2) and one E-SQUID device (E-SQ3). A back gate can be used to tune the global electronic properties of the devices, and measurements are performed in a dilution refrigerator with a base temperature of 40~mK.

Fig~\ref{Fig1}e shows a V-I curve for C-SQ1 at 40~mK, which displays a distinct supercurrent branch (inset shows the measurement configuration). In order to establish the presence of Josephson coupling we test whether the devices show clear SQUID behavior. We apply a current bias ($I$) close to the critical current ($I_c$) and monitor the voltage drop ($V$) as a function of perpendicular magnetic field ($B$). Fig~\ref{Fig1}f shows that all devices undergo periodic oscillations in $B$. Since we expect these oscillations to be periodic in the flux threading through the SQUID, a reduction in the loop area should result in a larger period in $B$. This is precisely what we observe when we compare C-SQ1 (upper panel) and C-SQ2 (middle panel), where C-SQ2 is designed to have a smaller loop area. E-SQ3 also shows similar periodic oscillations (lower panel) when the top gates are appropriately tuned (discussed in more detail below). The period ($\Delta B$) for each of the traces in Fig~\ref{Fig1}f can be determined by Fourier analysis (Fig~\ref{Fig1}g) to be 19~$\mu$T, 31~$\mu$T and 21~$\mu$T for C-SQ1, C-SQ2 and E-SQ3 respectively. This gives us an effective loop area $A_{eff} = \Phi_0/\Delta B$ ($\Phi_0 = h/2e$ is the flux quantum), which is consistently larger than the lithographically defined central (insulating) area. The difference arises from the Meissner effect in the superconducting areas, which focuses the applied field into center of the SQUID loop. We confirm this by using numerical simulations of 2D current distributions in thin film superconductors~\cite{Khapaev_Sim_SST} with an appropriate choice of the screening length (see SI for details). Taking into account the full geometry of the devices, we find the calculated period (see circles in Fig~\ref{Fig1}g) to agree well with the experiments.

\begin{figure}[!t]
\includegraphics[width=1\linewidth]{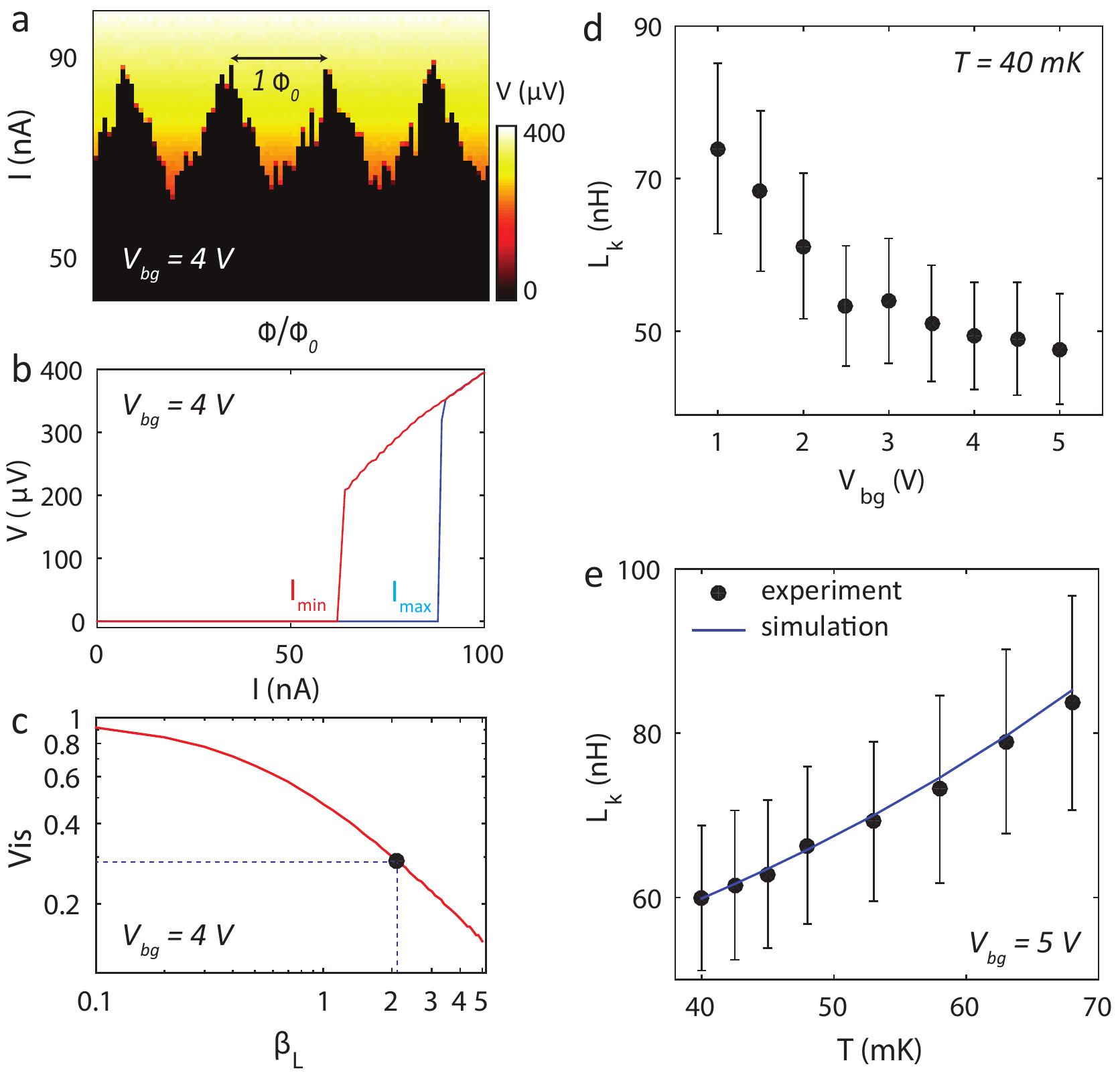}
\caption{\textbf{C-SQUID: back gate and temperature dependence} (a) Color maps of V-I curves for different values of normalized flux ($\Phi/\Phi_0$) at $V_{bg} = 4$~V. (b) Individual V-I traces from (a) showing the maximum ($I_{max}$) and minimum ($I_{min}$) critical currents. (c) Numerically simulated curve for $\textrm{\emph{Vis}}$ vs. $\beta_L$ in the noise free case with experimentally obtained value of $\textrm{\emph{Vis}}$ at $V_{bg} = 4$~V. (d) Variation of kinetic inductance ($L_k$) with $V_{bg}$. (e) Variation of $L_k$ with temperature at $V_{bg} = 5$~V (black circles) and comparison with numerical simulations (blue curve). These measurements were performed in the different cooldown, resulting in a slightly different value of $L_k$ as compared to (d).}
\label{Fig2}
\end{figure}

While the $V(\Phi)$ oscillations clearly demonstrate the successful creation of SQUIDs at the LAO/STO interface, analysis of $I_c$ ($\Phi$) oscillations provides a more quantitative understanding of the factors which determine the SQUID response. In the absence of thermal fluctuations the maximum critical current ($I_{max}$) across the SQUID is set by the Josephson coupling energy and the minimum critical current ($I_{min}$) is determined by the screening parameter $\beta_L = I_{max} L/\Phi_0$ ($L$ is the total inductance of the SQUID loop). In other words $L$ plays a crucial role in determining the visibility ($\textrm{\emph{Vis}} = \frac{I_{max}-I_{min}}{I_{max}}$) of the $I_c(\Phi)$ oscillations. The exact relation between $\textrm{\emph{Vis}}$ and $\beta_L$ can be obtained by numerical simulations (red curve in Fig.~\ref{Fig2}c). In order to experimentally determine $\textrm{\emph{Vis}}$, we keep the gate voltage fixed and record V-I's for different $\Phi$ (Fig.~\ref{Fig2}a) to estimate $I_{max}$ and $I_{min}$ (Fig.~\ref{Fig2}b). We find that for $V_{bg}=4$~V, $I_{max}=88$~nA and $\textrm{\emph{Vis}}\sim 0.3$. From Fig.~\ref{Fig2}c we estimate $\beta_L\sim 2.1$, giving $L\sim 50$~nH. This is nearly three orders of magnitude larger than the estimated geometric inductance of the SQUID loop. This additional inductance of the superconductor arises from the kinetic energy stored in the Cooper pairs and is known as the kinetic inductance ($L_k$). In general it can be expressed as $L_k\propto\frac{m^*}{n_sd}$, where $m^*$ is the effective mass of the charge carriers, $n_s$ is the superfluid density and $d$ is the thickness of the superconductor. The 2D nature of the LAO/STO interface ($d\sim 10$~nm)~\cite{Reyren_Aniso_APL}, combined with an extremely low $n_s$~\cite{Bert_Moler_superfluid_PRB} and large $m^*$~\cite{McCollam_Meff_APLMat} naturally result in a greatly enhanced kinetic inductance. While for most SQUID designs the contribution of $L_k$ can be neglected, here the SQUID response is in fact dominated by $L_k$. Thus, a straightforward analysis of the $I_c(\Phi)$ oscillations, as described above, allows us to directly estimate $L_k$ of our SQUID loop.

Fig.~\ref{Fig2}d shows that the kinetic inductance of the SQUID can be continuously tuned with the back gate (see SI for details of the analysis and error estimates). To our knowledge this is the only intrinsic superconductor where the kinetic inductance can be tuned in-situ via the field effect. Increasing $V_{bg}$ induces more carriers at the LAO/STO interface which in turn increases the superfluid density~\cite{Bert_Moler_superfluid_PRB}. This results in an overall decrease in $L_k$. In addition to the back gate, we expect the temperature to also have a substantial effect on $L_k$. Increasing the temperature should reduce $n_s$, thereby increasing $L_k$. Indeed, Fig.~\ref{Fig2}e clearly shows that $L_k$ increases with temperature. We therefore find that both the back gate and temperature dependence of $L_k$ are mutually consistent with the picture that the SQUID modulations are determined predominantly by the kinetic inductance. We compare our results in Fig.~\ref{Fig2}e with numerical simulations, solving the London equations for our geometry~\cite{Khapaev_Sim_SST}. Using the Ginzburg-Landau expression for the London penetration depth $\lambda_L(T) = \lambda_L(0)/(1-T/T_c)^{1/2}$~\cite{Tinkham_Intro} and the experimentally determined $T_c = 213$~mK, we find a good agreement between the experiments and simulations (blue curve). For the Pearl length $\Lambda_p(T) = 2\lambda_L(T)^2/d$ we obtain a value of 3~mm at $T=40$~mK, which is similar in magnitude to the value obtained via scanning SQUID measurements~\cite{Bert_Moler_superfluid_PRB}. Furthermore, we find that $L_k$ is dominated by the constrictions, which behave as quasi-1D structures connected to the 2D bulk superconducting reservoirs (see SI for a full description of the calculations).

\begin{figure*}[!t]
\includegraphics[width=1\linewidth]{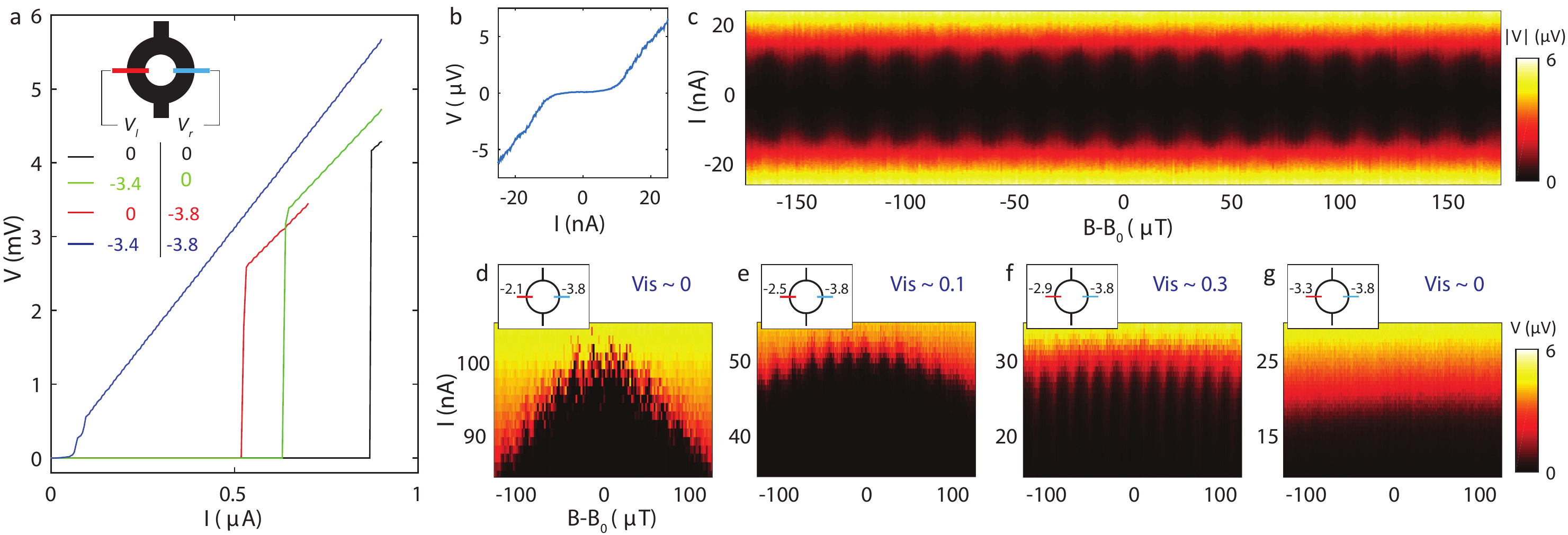}
\caption{\textbf{E-SQUID: controllable Josephson junctions} (a) V-I curves for different combinations of left gate voltage ($V_l$) and right gate voltage ($V_r$), with $V_{bg}=-1$~V. Inset shows a schematic of the device. (b) Zoom in of the blue trace in (a). (c) 2D plot showing SQUID oscillations with the top gates optimally tuned ($V_l = -3.4$~V, $V_r = -3.8$~V). (d)-(g) Variation in the visibility of the modulations as $V_l$ is reduced, with $V_r$ held constant. The values of $V_l$ and $V_r$ are indicated in the left inset. These measurements were performed in a different cooldown, resulting in slightly different values of the gate voltages as compared to (a)-(c).}
\label{Fig4}
\end{figure*}


The C-SQUIDs describe a simple, yet effective way to demonstrate quantum interference at the LAO/STO interface. However, they do not allow one to locally control the weak links. In contrast the E-SQUID (described earlier, see Fig.~\ref{Fig1}c,d) uses local top gates to create and control JJs in each arm of the SQUID. Though some previous studies with a single top gate gate have shown evidence of a gate controllable Josephson effect~\cite{Goldhaber_JJ,Goswami_NL,Bal_JJ_APL}, there was no clear observation of quantum interference. We now describe the operation of the E-SQUID in detail. When no gate voltages are applied to either of the gates ($V_l = V_r = 0$~V) the device is equivalent to a simple superconducting loop. The black trace in Fig.~\ref{Fig4}a shows the corresponding V-I trace. When a large negative voltage is applied only to the left gate $V_l = -3.4$~V the current flow through the left arm decreases, thereby reducing the total critical current across the loop (green trace). The red trace shows a similar V-I curve when only the right gate is made highly negative ($V_r = -3.8$~V) (refer to the SI for more details about the tuning procedure). As expected, none of these three electrostatic configurations produce SQUID oscillations. However, when both gates are depleted ($V_l = -3.4$~V and $V_r = -3.8$~V, blue trace) the critical current reduces significantly. Fig.~\ref{Fig4}b shows the V-I trace in this gate configuration over a smaller range. At this point we observe distinct SQUID oscillations (Fig.~\ref{Fig4}c), thereby demonstrating the existence of an electrostatically defined JJ in each of the arms. We note that this process is completely reversible, whereby removing the gate voltages brings the device back to its original state, with no JJs.

The sensitivity of the JJs to the top gate voltages defines an optimal operating range for the E-SQUID. To quantify this we keep the right gate fixed at $V_r = -3.8$~V and monitor the visibility for different values of $V_l$, as shown in Fig.~\ref{Fig4}d-g. The left inset shows the voltages applied to the left/right gates (these measurements were performed in a different cooldown to those described in Fig.~\ref{Fig4}a-c, and therefore the absolute values of the voltages are somewhat different). When the left gate is relatively open (Fig.~\ref{Fig4}d) the oscillations are hardly visible ($\textrm{\emph{Vis}}\sim 0$). As $V_l$ is made more negative the visibility increases, reaching a maximum value of 0.3 (Fig.~\ref{Fig4}f). By depleting the region below the left gate even further, the oscillations disappear again. This continuous transition can be physically understood as follows. Since the top gates act locally, their influence on the superconducting banks is minimal. Therefore the maximum visibility is obtained when both JJs have the same critical current. This condition is satisfied for Fig.~\ref{Fig4}f ($V_r = -3.8$~V, $V_l = -2.9$~V). Tuning $V_l$ away from this optimal condition thus increases the asymmetry between the two JJs resulting in a reduced visibility.

\begin{figure}[!t]
\includegraphics[width=1\linewidth]{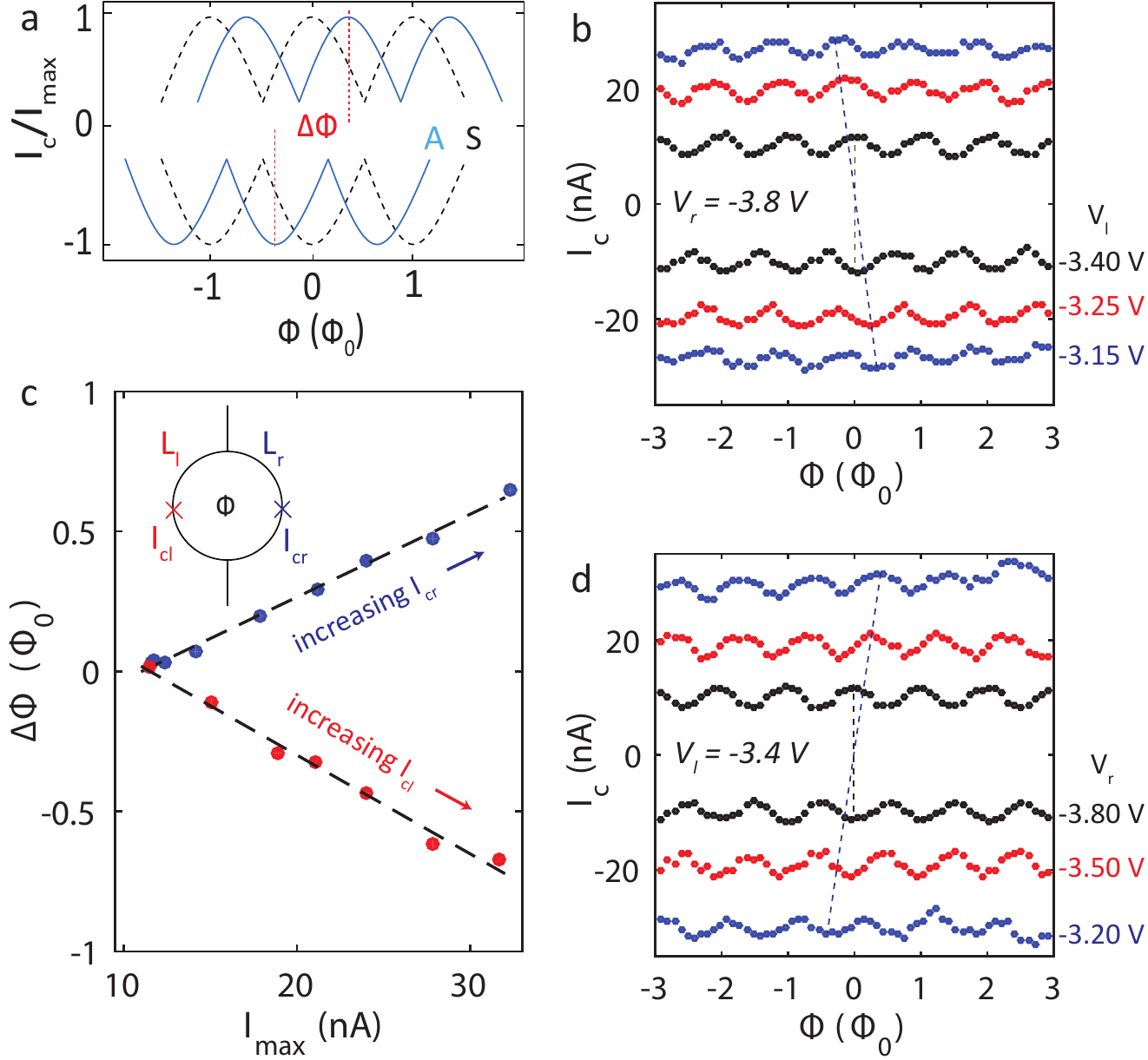}
\caption{\textbf{E-SQUIDs: controllable asymmetry} (a) $I_c$ ($\Phi$) oscillations for a symmetric ($S$) SQUID (black dashed curve) and an asymmetric ($A$) SQUID with large kinetic inductance (blue solid curve). The maximum and minimum values of $I_c$ shift along the $\Phi$-axis by an amount $\Delta\Phi$. (b) $I_c (\Phi)$ oscillations with $V_r$ fixed and selected values of $V_l$. Similar plots, but for fixed $V_l$ and varying $V_r$ are shown in (d). (c) Variation of $\Delta\Phi$ with $I_{max}$ extracted from (b),(d) but also including more values of top gate voltage. Dashed lines show linear fits. Inset shows a schematic of an asymmetric SQUID.}
\label{Fig5}
\end{figure}

For SQUIDs with a small loop inductance the dominant source of asymmetry arises due to unequal values of critical current in the two JJs (i.e., $I_{cl}\neq I_{cr}$, see SQUID schematic in inset to Fig.~\ref{Fig5}c). On the other hand, when $L_k$ is large (as is the case for LAO/STO) one must consider the combined effects of asymmetries in $I_c$ and $L_k$ in the two arms of the SQUID loop (see Ref~\cite{Squid_Handbook} for a description about asymmetric SQUIDs). The most important consequence of such asymmetry is that $I_c (\Phi)$ curves are offset along the $\Phi$-axis (Fig.~\ref{Fig5}a). Such offsets arise due to the large $L_k$, which produces a substantial self-flux ($\Phi_s$), in addition to the applied flux $\Phi$. When the phase drop across each JJ reaches $\pi/2$, $I_c$ reaches its maximal value $I_{max} = I_{cl}+I_{cr}$ and $\Phi_s(+) = I_{cr}L_r - I_{cl}L_l$. Reversing the direction of current bias results in the same magnitude of self flux, but now of the opposite sign (i.e., $\Phi_s(-) = -\Phi_s(+)$.) Thus $\Delta\Phi = 2(I_{cr}L_r - I_{cl}L_l)$, where $\Delta\Phi = \Phi_s(+)-\Phi_s(-)$.

By controlling the two JJs in our E-SQUID we study the effects of such asymmetry in the SQUID response. In particular, we show that the ability to independently tune the critical current of each JJ gives us an alternative method to extract the kinetic inductance. We start with an electrostatic configuration identical to the one in Fig.~\ref{Fig4}c and plot $I_c (\Phi)$ (black curve in Fig.~\ref{Fig5}b). The black dashed line confirms that there are no discernible offsets in the $\Phi$-axis and the SQUID is in a symmetric configuration. We now hold $V_r$ constant (i.e, $I_{cr}$ does not change) and make $V_l$ less negative (increase $I_{cl}$). We find that the $I_c (\Phi)$ curves move towards the left (right) for positive (negative) current bias. The blue dashed line clearly indicates that $\Delta\Phi$ acquires a negative sign. Performing the same experiment with $V_l$ fixed and opening the right gate, we expect $I_{cr}$ to increase, thereby inducing a self-flux in the opposite direction. This sign reversal of $\Delta\Phi$ can be seen in Fig.~\ref{Fig5}d.

The variation of $\Delta\Phi$ with $I_{max}$ is plotted in Fig.~\ref{Fig5}c. The blue (red) points correspond to measurements performed with $V_r$ ($V_l$) varying while the other gate is fixed. Since $\frac{\partial\Delta\Phi}{\partial I_{cr}} = 2L_r$ and $-\frac{\partial\Delta\Phi}{\partial I_{cl}} = 2L_l$, linear fits to these points (dashed lines) allow us to estimate $L_r \sim 31$~nH and $L_l \sim 36$~nH. This difference is within the error bars of our estimates and we conclude that any intrinsic asymmetry in $L_k$ of the two arms is small. Thus, the observed shifts along the $\Phi$-axis arise from a combination of the large $L_k$ and unequal critical currents of the JJs. This is a particularly important finding in the context of LAO/STO since it suggests that any mesoscopic inhomogeneities in the superfluid density~\cite{Honig_Ilani_Nat_Mat,Kalisky_Moler_Nat_Mat} average out over a length scale of a few microns and do not have a considerable effect on the operation of these SQUIDs. Furthermore, even if such inhomogeneities are present, using the E-SQUID it is always possible to appropriately tune the critical current of the JJs to minimize the effects of self-flux.

The ability to probe the phase of the superconducting order parameter naturally opens the door to answer more specific questions about the pairing symmetry. To do so, one could combine the LAO/STO superconductor with a standard $s$-wave superconductor to create devices analogous to the corner SQUIDs/JJs in high $T_c$ cuprates~\cite{Harlingen_RMP}. From a technological perspective, our studies of the E-SQUID demonstrate a completely new variety of JJs which are both electrostatically defined and electrostatically controlled. Such an architecture to create JJs eliminates any detrimental effects of physical interfaces between dissimilar materials. Detailed transport spectroscopy studies should allow one to ascertain whether such electrostatic interfaces are in fact superior.

\textbf{Acknowledgements:} We thank Teun Klapwijk, Attila Geresdi, Anton Akhmerov, Alexander Brinkman, and Jochen Mannhart for useful discussions and feedback about the preliminary results. This work was supported by The Netherlands Organisation for Scientific Research (NWO/OCW) as part of the Frontiers of Nanoscience program, the Dutch Foundation for Fundamental Research on Matter (FOM), the Deutsche Forschungsgemeinschaft (DFG) via Project KO 1303/13-1 and by the EU-FP6-COST Action MP1308.

\textbf{Author Contributions:} E.M. fabricated the devices. S.G. performed the transport measurements with help from E.M. S.G. and A.M.R.V.L.M. analyzed the data. R.W., R.K. and D.K. carried out the numerical simulations and Y.M.B. provided theoretical support. L.M.K.V. and A.D.C. supervised the project. S.G. wrote the manuscript with inputs from all co-authors.

\clearpage
\setcounter{figure}{0}
\renewcommand{\thefigure}{S\arabic{figure}}
\onecolumngrid
\section{\large{Supplementary Information}}
\vspace{10mm}

\begin{section}{Fabrication}

Fig.~S1a shows the complete flow of the fabrication processes. Single crystal TiO$_2$-terminated (001) SrTiO$_3$ (STO) substrates are purchased from CrysTec GmbH$^\copyright$ and used without any modification. As discussed in the main text, we fabricate two varieties of SQUIDs: C-SQUID (red arrow) and E-SQUID (brown arrow). Fabrication of the E-SQUID starts with electron beam lithography (EBL) and subsequent deposition of tungsten (W) markers and lift-off. This step is not necessary for the C-SQUID since it requires only a single EBL step. After W deposition (only for the E-SQUID), samples are processed with EBL followed by deposition of amorphous LaAlO$_3$ (a-LAO) via pulsed laser deposition (PLD) to define insulating regions. Subsequent lift-off and optical inspection shows these masked regions (Fig.~S1b). Then, crystalline LaAlO$_3$ (c-LAO) is grown via PLD monitored in-situ by reflection high energy electron diffraction (RHEED) to confirm layer-by-layer growth (Fig.~S1c). The RHEED diffraction pattern also confirms the two-dimensional nature of the growth (inset of Fig.~S1c). Growth of c-LAO is the last fabrication step for C-SQUID. For the E-SQUID, we perform another EBL to define the top gates, followed by metal (Au) evaporation and lift-off. Optical images of a final E-SQUID device are shown in Fig.~S1d,e. Ultrasonic wedge bonding is used to make electrical contact to the conducting regions. Further details about each step of the fabrication have been reported previously~\cite{Goswami_NL_Supp}. We point out two important differences relevant for the devices reported in this study: (i) the c-LAO was grown at a higher temperature of 840$^\circ$C resulting in a larger critical temperature of the superconductor and (ii) the metallization for the top gates was preceded by a short oxygen plasma exposure to remove polymer residues.

\begin{figure*}[!b]
\includegraphics[width=.9\linewidth]{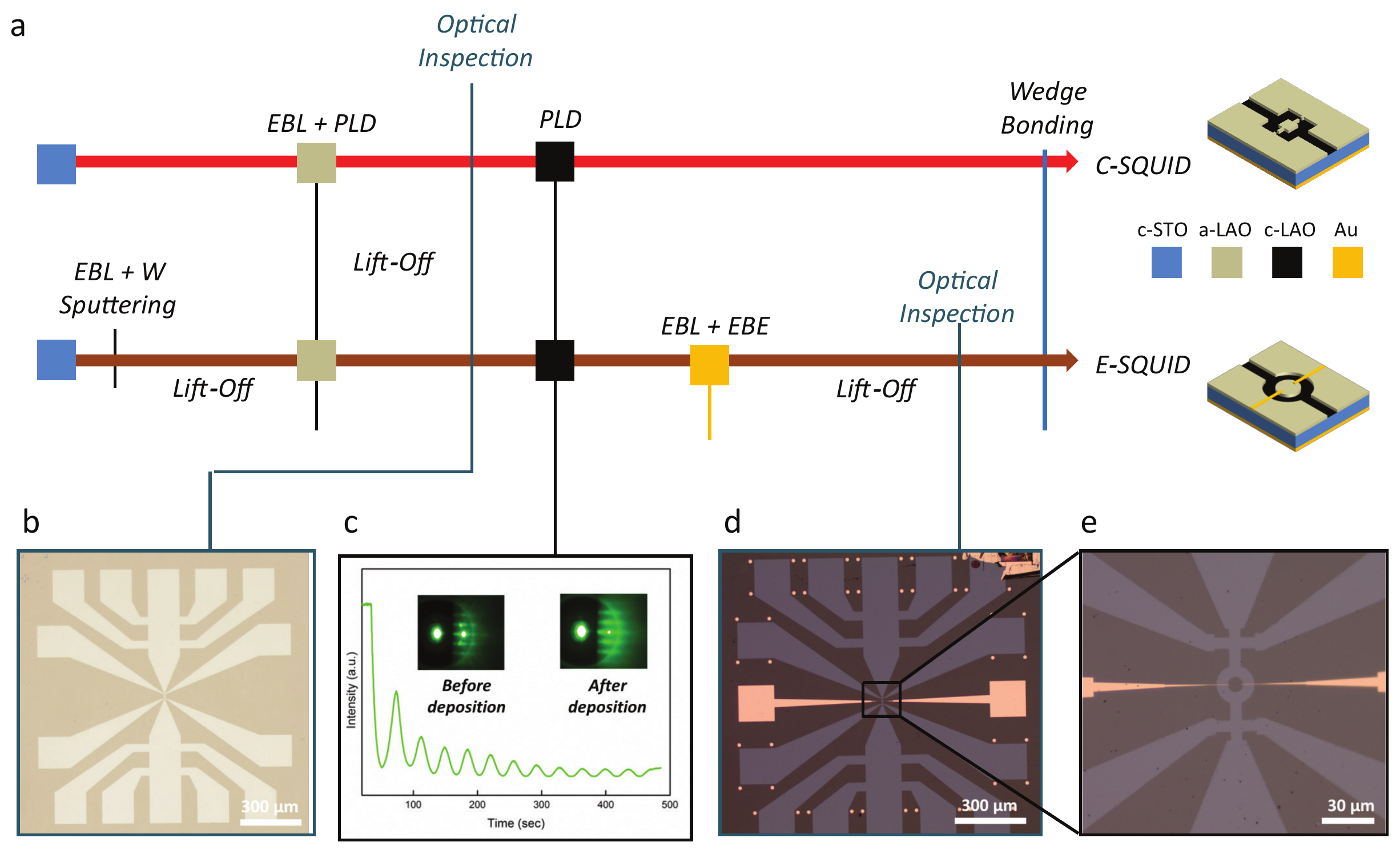}
\caption{(a) Fabrication flow for SQUID devices. (b) Optical image after deposition of the amorphous LAO mask. (c) RHEED oscillations obtained during the growth of c-LAO. Inset shows diffraction patterns before/after the growth. (d) Optical image after top gates deposition. (e) Zoom in of (d) showing the active area of a E-SQUID device.}
\label{fab}
\end{figure*}

\end{section}

\newpage
\begin{section}{Error estimates for kinetic inductance}

To estimate the error bars for $L_k$ (see Fig.~2d of main text) we analyze the $I_c (\Phi)$ oscillations for each value of $V_{bg}$. We extract average values of $I_{max}$ and $I_{max}-I_{min}$ (Fig.~S2a,b) from 5-6 consecutive oscillations. The error bars represent the mean absolute deviation and Fig.~S2c shows the resulting error in \emph{Vis}. We see that \emph{Vis} does not change with $V_{bg}$ and we obtain  $\textrm{\emph{Vis}}\sim 0.29\pm 0.03$. Since \emph{Vis} is a strong function of $\beta_L$ and $L_k = \beta_L\Phi_0/I_{max}$, this uncertainty in \emph{Vis} results in relatively large errors ($\sim 15\%$) on $L_k$.

\begin{figure*}[!t]
\includegraphics[width=.9\linewidth]{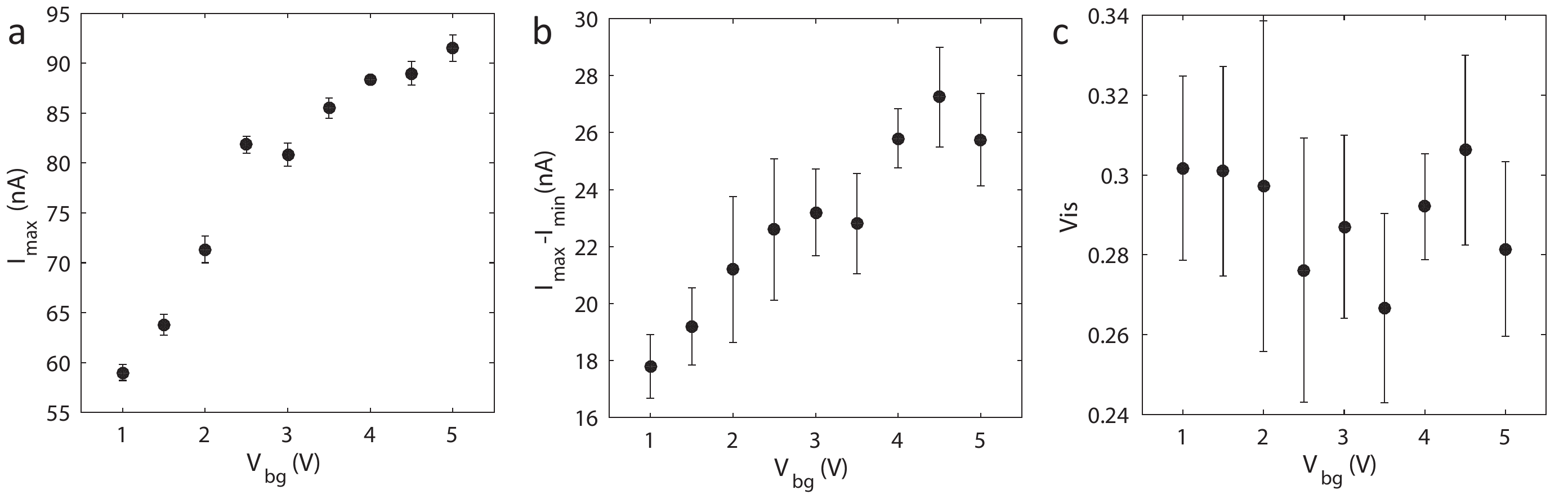}
\caption{Error estimates for (a) the maximum critical current $I_{max}$, (b) $I_{max}-I_{min}$, and (c) visibility for different values of back gate voltage.}
\label{err}
\end{figure*}

\end{section}

\begin{section}{Tuning the E-SQUID}

To tune the E-SQUID into the optimal configuration we start by first recording V-Is as each top gate is depleted (with the other grounded). Fig.~S3a(b) show such plots for $V_{rg} = 0$~V ($V_{lg} = 0$~V), where the dark/bright regions indicate the superconducting ($S$)/resistive ($R$) branch of the V-I curve and the border is representative of the critical current. These plots give us an indication of the action of the individual gates, and allow us to identify values of gate voltage for which the regions below the top gates are being actively depleted ($I_c$ decreases significantly). Next, we keep one gate fixed at a large negative voltage (Fig.~S3c) and vary the other. Below a certain gate voltage (indicated by the dashed line) SQUID oscillations begin to appear. To check for the presence of these oscillations we found it particularly convenient to look for $V(\Phi)$ oscillations (as shown in Fig.~1f of the main text), since such measurements are much quicker than $I_c (\Phi)$ measurements.
\begin{figure*}[!b]
\includegraphics[width=.9\linewidth]{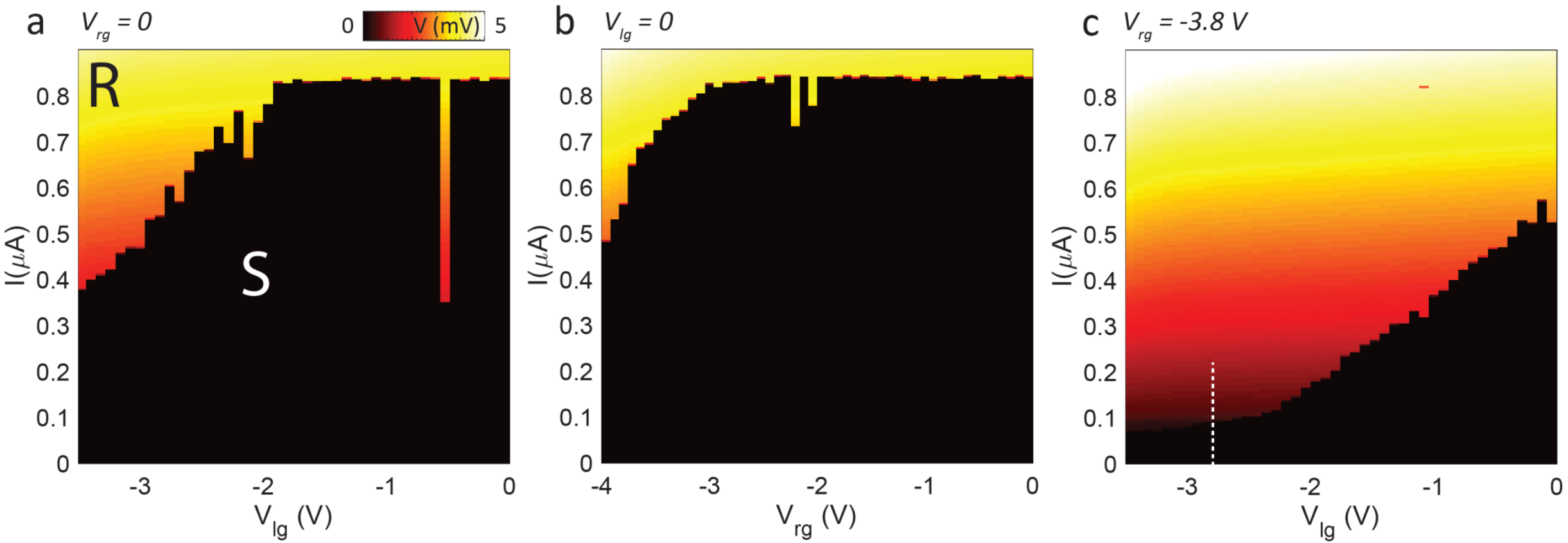}
\caption{2D color maps showing the variation of critical current with (a) $V_{lg}$ varied, $V_{rg} = 0$~V, (b) $V_{rg}$ varied, $V_{lg} = 0$~V, and (c) $V_{lg}$ varied, $V_{rg} = -3.8$~V.}
\label{tuning}
\end{figure*}

\end{section}

\newpage
\begin{section}{Finite Element Simulations}

\begin{figure*}[!b]
\includegraphics[width=1\linewidth]{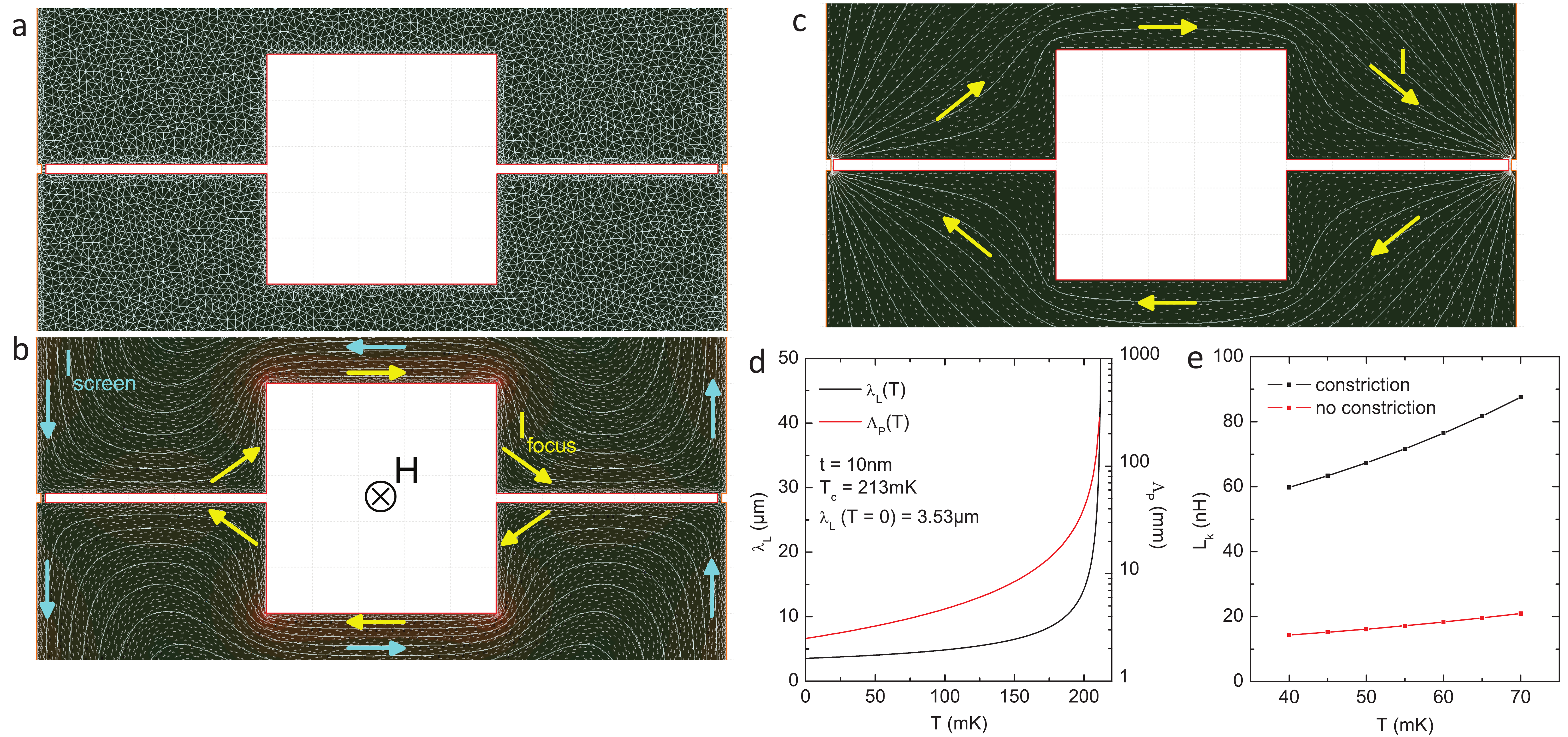}
\caption{(a) Example of a mesh used for finite element simulations. Results of numerical simulations for (b) effective area, and (c) kinetic inductance. See text for more details. (d) Calculated temperature dependence of London penetration depth and Pearl length. (e) Calculated temperature dependence of kinetic inductance, with and without constrictions.}
\label{fes}
\end{figure*}

For the simulations of effective area $A_{\mathrm{eff}}$ and kinetic inductance $L_{\mathrm{k}}$ we use the software package 3D-MLSI~\cite{Khapaev_Sim_SST_Supp}. A brief summary of the theoretical background is presented in this section. For more details we refer to~\cite{Khapaev_Sim_SST_Supp,Brandt_PRB} and references therein.

Initially, a given superconducting circuitry with thickness $t$ extended in the $x$-$y$-plane is subdivided into finite sized triangles (finite element method); cf. Fig.~S4a). For each individual element, static London equations are applied to compute the current distribution and the associated magnetic fields for the overall structure under appropriate global boundary conditions. Here, the central figure of merit is the thickness-integrated current density

\begin{equation}
\vec{J}(x,y) = \int \vec{j}(x,y,z) dz= (J_x, J_y)
\label{eq:sheet_current}
\end{equation}

labeled sheet current. As long as $t$ is much smaller than all other dimensions of the superconducting film, the assumption $\nabla \cdot \vec{J}(x,y) = 0$ remains valid and the sheet current can be expressed by a scalar potential $g$

\begin{equation}
\vec{J}(x,y) = -\widehat{z} \times \nabla g = (\frac{\partial g}{\partial y}, -\frac{\partial g}{\partial x})
\label{eq:stream_function}
\end{equation}

also called stream function. $g(x,y)$ is calculated for each finite element with $g(x_1,y_1)-g(x_2,y_2)$ being the current crossing any line connecting $(x_1,y_1)$ and $(x_2,y_2)$. Furthermore, $g(x,y)$ can be used to calculate the full energy functional $E$ of the thin film which in turn is closely related to the inductance matrix $L$ comprising self and mutual inductances. By this means, also fluxoids $\Phi$ can be calculated for superconducting films containing holes or slits.

For the calculation of effective areas $A_{\mathrm{eff}}$ a homogeneous external magnetic field $H = 1\,\mathrm{mA/\mu m}$ is applied perpendicular to the plane of the SQUID. Boundary conditions are chosen to be $g(x,y) = 0$ at the edge of the SQUID hole as well as at the outer edge of the superconducting structure, i.e. the net current circulating around the hole is $I = 0$. In other words, the current $I_{\mathrm{screen}}$ flowing close to the outer edge of the superconductor screens the applied field into the periphery of the SQUID while the current $I_{\mathrm{focus}}$ flowing at the edge of the hole focuses $H$ into the SQUID loop. $I_{\mathrm{screen}}$ and $I_{\mathrm{focus}}$ flow in counter-clockwise and clockwise direction since the net current is set to zero. $A_{\mathrm{eff}}$ is obtained by computing the fluxoid in the hole and by $A_{\mathrm{eff}} = \Phi/({\mu_0 H})$, with the vacuum permeability $\mu_0$. Fig.~S4b) displays the current distribution and $g(x,y)$ for the calculation of $A_{\mathrm{eff}}$ for device C-SQ1. Here, white solid lines represent the contour lines of $g(x,y)$. Short white lines indicate $\vec{J}(x,y)$. For a better understanding, $I_{\mathrm{focus}}$ and $I_{\mathrm{screen}}$ have been emphasized by yellow and cyan arrows, respectively. The color code gives the amplitude of local currents with red and black areas indicating high and low current amplitude, respectively.

Turning to the calculation of the kinetic inductance $L_k$, the boundary conditions are $g(x,y) = 1\,\mathrm{mA}$ at the edge of the SQUID hole and $g(x,y) = 0$ at the outer edge of the superconductor, i.e. $L_k$ is computed for a current with amplitude $I = 1\,\mathrm{mA}$ circulating around the hole; cf. Fig.~S4c). Again, the circulating current $I$ has been emphasized by yellow arrows. Film thickness $t$ and London penetration depth $\lambda_{\mathrm{L}}$ are crucial input parameters for the calculation of $L_k$, however these quantities are initially unknown. We assume $t = 10\,\mathrm{nm}$ and subsequently adjust $\lambda_{\mathrm{L}}$ to reproduce the measured $L$ from the experiments. Choosing $\lambda_{\mathrm{L}} = 3.9\,\mu m$, $T = 40\,\mathrm{mK}$ and $T_{\mathrm{c}} = 213\,\mathrm{mK}$, we find good agreement between simulation and measurement. We note here, that these values give a Pearl length of $\Lambda_{\mathrm{P}} = 2 \lambda_{\mathrm{L}}^2/t = 3.04\,\mathrm{mm}$ at $T = 40\,\mathrm{mK}$ which is of the same order of magnitude as reported elsewhere~\cite{Bert_Moler_superfluid_PRB_Supp}.

To reproduce the functional evolution of $L_k(T)$ as seen in the experiments, we implement a temperature dependence of $\lambda_{\mathrm{L}}$. We find reasonable agreement for

\begin{equation}
\lambda_{\mathrm{L}}(T) = \frac{\lambda_{\mathrm{L}}(T = 0)}{\sqrt{1-T/T_{\mathrm{c}}}}\textrm{ ,}
\label{eq:lambda_L}
\end{equation}

which is plotted together with the related $\Lambda_{\mathrm{P}}(T)$ in Fig.~S4d). Note that the same parameters ($\lambda_{\mathrm{L}}(T)$, $t$, $T_{\mathrm{c}}$) were used to calculate the effective area at $T = 40\,\mathrm{mK}$.

To stress the fact that for device C-SQ1 the dominant contribution to $L_k$ arises from the constriction type Josepshon junctions, we compare the behavior of $L_k(T)$ of C-SQ1 with an otherwise equivalent device without constrictions in Fig.~S4e. We find that the overall inductance is indeed governed by the contribution of the constrictions by at least a factor of 4. This can be attributed to a large kinetic inductance since the dimensions of the constrictions are well below $\lambda_{\mathrm{L}}$.

\end{section}

\begin{section}{Comparison with the RCSJ model}

\begin{figure*}[!b]
\includegraphics[width=0.8\linewidth]{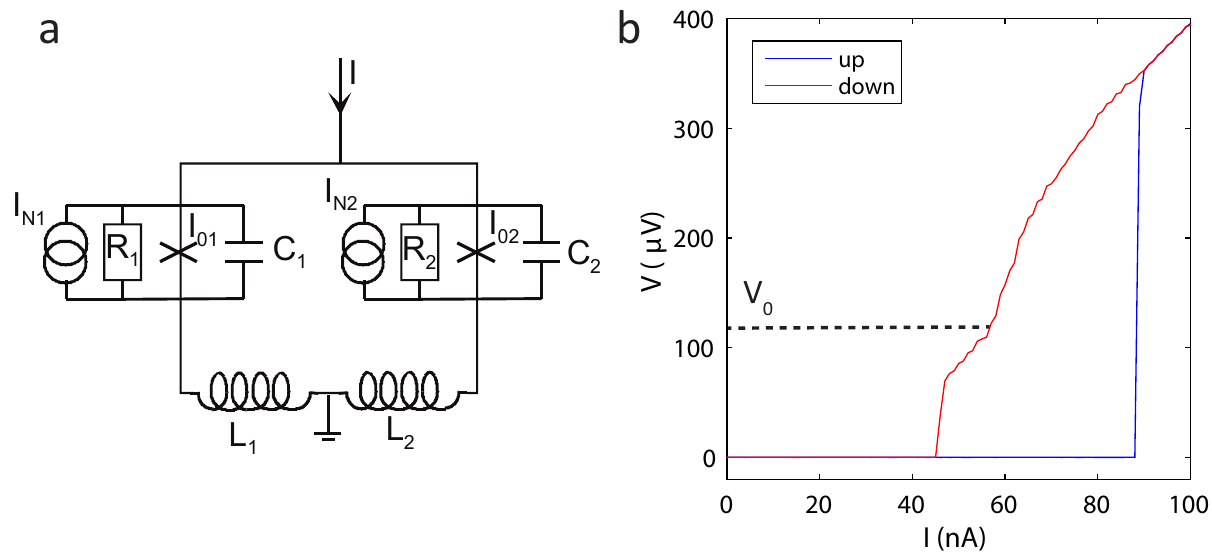}
\caption{(a) Circuit for RCSJ model. (b) Experimental V-I trace at $V_{bg} = 4$~V, showing hysteresis. $V_0$ indicates the position of an LC resonance.}
\label{RCSJ}
\end{figure*}

In our experiments we found that the nature of the V-I curves (value of critical current, shape of the transition to the resistive branch) depends on the back gate voltage. Since our estimates of $L_k$ rely on the conversion of visibility (\emph{Vis}) to $\beta_L$, we implicitly assume that the resistively and capacitively shunted junction (RCSJ) model~\cite{Stewart_APL,McCumber_JAP} appropriately describes our system. Below, we briefly describe the RSCJ model and how it relates to our C-SQUID devices.

To model an asymmetric SQUID we consider the circuit shown in Fig.~S5a. The Josephson junctions are described by Josephson currents with phases $\delta_1$  and $\delta_2$  and amplitudes $I_{01} = I_0(1-a_i)$  and $I_{02} = I_0(1+a_i)$ , resistors $R_{1} = R/(1-a_r)$ and $R_{2} = R/(1+a_r)$ , and capacitors $C_{1} = C(1-a_c)$  and $C_{1} = C(1+a_c)$ . The Nyquist noise arising from the two resistors is described by two independent current noise sources $I_{N1}$ and $I_{N2}$ having white spectral power densities $4k_BT/R_1$ and $4k_BT/R_2$, respectively and $k_B$ is the Boltzmann constant. The two arms of the SQUID loop have inductances $L_{1} = L(1-a_l)/2$ and $L_{1} = L(1+a_l)/2$. The total inductance L is the sum of the geometric ($L_g$) and the kinetic ($L_k$) inductance. The loop is biased with a current $I$. A flux $\Phi_{ext}$ is applied to the loop. In the following we normalize currents to $I_0$, fluxes to the flux quantum $\Phi_0$, voltages to $I_0R$ and time to $\Phi_0/2\pi I_0R$ . We further use the McCumber parameter $\beta_c = 2\pi I_0R^2C/\Phi_0$, the inductance parameter $\beta_L = 2I_0L/\Phi_0$  and the noise parameter $\Gamma = 2\pi k_BT/(I_0\Phi_0)$. With these units the SQUID is described by the differential equations~\cite{Tesche_JLTP,Bruines_JLTP}

\begin{equation}
i/2+j = \beta_c(1-a_c)\ddot{\delta_1}+(1-a_r)\dot{\delta_1}+(1-a_i)\sin\delta_1 + i_{N1}
\end{equation}

\begin{equation}
i/2-j = \beta_c(1+a_c)\ddot{\delta_2}+(1+a_r)\dot{\delta_2}+(1+a_i)\sin\delta_2 + i_{N2}\textrm{ .}
\end{equation}

Here $i$ and $j$ denote the normalized bias current and circulating current in the SQUID loop and the dots denote derivatives with respect to normalized time. The spectral densities of the normalized noise currents $i_{N1}$ and $i_{N2}$ are given by $4\Gamma (1-a_r)$ and $4\Gamma (1+a_r)$ , respectively. The circulating current $j$ is given by

\begin{equation}
j = \frac{1}{\beta_L}\left(\frac{\delta_2-\delta_1}{\pi}-2\Phi_{ext}+a_l\beta_L\frac{i}{2}\right)\textrm{ .}
\end{equation}

From these equations we get the time dependent (high frequency) voltages across the junctions via the second Josephson relation, $\dot{\delta_k} = u_k$, (k = 1,2). The dc (i.e. low frequency) voltage $v$ across the SQUID is obtained by averaging $(u_1+u_2)/2$ over sufficiently long times. From the dc voltage $v$ one obtains current voltage (IV) characteristics $i$ vs. $v$, the SQUID voltage modulation $v$ vs. $\Phi_{ext}$, the transfer function $dv/d\Phi_{ext}$ or the modulation of the critical current $i_c$ vs. $\Phi_{ext}$ and $\textrm{\emph{Vis}} =(i_{max}-i_{min})/i_{max}$. For the simulations discussed here we assumed symmetric parameters, i.e. $a_i=a_r=a_c=a_l=0$. Calculations for $\Gamma=0$ are shown in Fig.~2c of main text.

For higher values of back gate voltage $V_{bg}$, $I_c$ is significantly larger showing a distinct switch in the V-I curves. Here we find a reasonable agreement with the RCSJ model. For example for $V_{bg}=4$~V (Fig.~2b of main text) numerical simulations give $\beta_L\sim 2.3$ and $\beta_c\sim 5.5$ for the noiseless case. Including thermal fluctuations ($\Gamma = 0.015$), we find an upper limit for $\beta_c\sim 15$. In either case, since $\beta_c>1$, we expect to observe hysteretic V-I curves. The experimental V-I trace (Fig.~S5b.) clearly displays such a hysteresis. Furthermore, we observe a distinct bump-like feature in the return branch of the curve (red trace). In the RCSJ framework, this structure arises due to an LC resonance from the circuit shown in Fig.~S5a. One would then expect that $eV_0 = \hbar\omega = 1/\sqrt{LC/2}$. Here $V_0\approx 120$~$\mu$V is the position of the feature and $\omega$ is the resonance frequency. Using the experimentally determined value of $L_k\approx 50$~nH we estimate $C\approx 1$~fF. Finally using $I_0=44$~nA and $R=8$~k$\Omega$ we obtain $\beta_c \approx 10$. This is indeed consistent with values of $\beta_c$ predicted by the RCSJ model. Therefore, this analysis shows that (a) the RCSJ model describes our system reasonably well and (b) our estimates of $L_k$ from the $I_c$ oscillations are indeed correct.

We found that the V-I curves at lower back gate voltage (as in Fig.~1e of the main text) could not be reproduced well within the RCSJ framework. In particular, we see a large excess current, which could be indicative of non-equilibrium processes or even a multi-valued current-phase relation. Furthermore, the low value of the critical current makes the effects of thermal noise much more significant, thus making it difficult to reliably extract the kinetic inductance. We therefore restrict our analysis to $V_{bg}\geq 1$~V (as seen in Fig.~2d of the main text).

\end{section}

\end{document}